\documentclass[multphys,vecphys]{svmult}

\usepackage{makeidx}     
\usepackage{graphicx}    
\usepackage{multicol}    

\makeindex             

\def\ale{\mathrel{\hbox{\rlap{\hbox{\lower4pt\hbox{$\sim$}}}\hbox{$<$}}}}
\def\age{\mathrel{\hbox{\rlap{\hbox{\lower4pt\hbox{$\sim$}}}\hbox{$>$}}}}

\begin{document}

\title*{Optical Bumps in Cosmological GRBs as Supernovae}

\titlerunning{Optical GRB Bumps as Supernovae} 
\author{J. S. Bloom\inst{1,2}}
\authorrunning{J. S. Bloom}

\institute{Harvard-Smithsonian Center for Astrophysics, MC 20, 
60 Garden Street, Cambridge, MA 02138, USA; {\tt{jbloom@cfa.harvard.edu}} \and Harvard Society of Fellows, 78 Mount Auburn Street, 
Cambridge, MA 02138, USA}

%
%

\maketitle

\begin{abstract}

From both photometric and broadband spectral monitoring of
$\gamma$-ray burst (GRB)\index{GRB}\index{Gamma-ray Burst} lightcurve
``bumps,'' particularly in GRB\,011121, a strong case grew for a
supernova (SN) origin. The GRB-SN connection was finally solidified
beyond a reasonable doubt with the discovery that the bump in
GRB\,030329\index{030329} was spectroscopically similar to a bright
Type Ic SN. In light of this result, I redress the previous SN bump
claims and conclude that 1) the distribution of GRB-SN bump peak
magnitudes is consistent with the local Type Ibc SNe peak distribution
and suggest that 2) the late-time bumps in all long-duration GRBs are
likely supernovae\index{supernova}.

\end{abstract}

\section{Introduction}

The connection of GRBs to supernovae (SNe) -- and hence the death of
massive stars -- was explored theoretically \cite{col68} even before
the discovery of GRBs. In the subsequent discovery paper, Klebesadel,
Strong, \& Olson reported on a catalog search for SNe coincident in
time and position with the first 13 GRBs known
\cite{kso73}.  In retrospect, the lack of a clear association is easy to 
explain: not only were SNe sparsely discovered, but the highest
redshift SN discovered by 1973 (SN\,1968P; $z = 0.11$) was less than
the redshift of all $\sim$35 classic GRBs redshifts known today. With
the development of a model for the production of a GRB from a
collapsing massive star \cite{woo93}, came renewed interest in the
possibility of a GRB--supernova connection. The modern form of and new
spins on Woosley's collapsar was presented in great detail by Andrew
MacFadyen (this meeting).

The discovery of the low redshift SN\,1988bw, a bright type Ic
associated with GRB\,980425 \cite{gvv+98} implied that massive stars
were capable of producing at least some incarnation of a GRB: the low
energy release of GRB\,980425\index{GRB 980425} compared with all
other known cosmological GRB energies suggests it may be only a
sub-class of GRBs \cite{bkf+98}.  Attempts to theoretically unify
GRB\,980425 with the majority of the long-duration cosmological ($z >
0.1$) bursts fall short of explaining all the observational
differences (see papers by Iwamoto, Patat this meeting).

\begin{figure}[bht]
\centering
\includegraphics[height=3.6in]{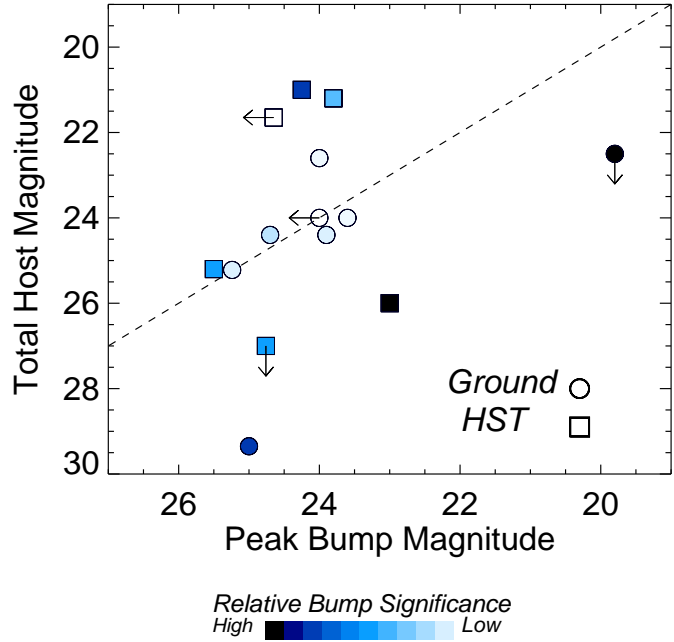}
\caption{The difficulty of detecting supernovae associated with GRBs: 
the magnitude of the host galaxy versus the peak magnitude of the
putative bump (in whichever optical filter the peak bump magnitude was
reported). Above the dashed line, bump magnitudes are fainter than the
integrated light of the host galaxy. The circles are measurements
where only ground-based imaging was used and squares are measurements
derived also using {\it Hubble Space Telescope} (HST)
\index{HST}\index{Hubble Space Telescope} imaging. 
Relative significances of the reliability of the bump detection were
assigned by the author based roughly on the number of detections of
the bump above the host light and the photometric error reported on
the bump peak. Updated (from Ref.\cite{blo03}) with new bumps in
030329 and 021211, reported in this meeting.}
\label{fig:1}       
\end{figure}

Our discovery of a faint red light curve ``bump'' superimposed on the
afterglow of GRB\,980326 \cite{bkd+99} was the first suggestion of a
SN associated with a cosmological GRB. Supernovae-like features were
later found photometrically in a number of bursts, some more plausibly
than others; the principal difficulty is that SNe peak at a
brightnesses comparable to their respective host galaxies (see
Fig.~\ref{fig:1}; Refs.~\cite{blo02,blo03}). Our multicycle effort on
HST (Kulkarni, PI) was designed to search for late-time afterglow
features and discriminate between other physical models for such
emission. While a few bursts showed no evidence for a SN (e.g.,
ref.~\cite{pks+03}), the late-time multicolor HST light curve and
broadband spectra of GRB\,011121 was reliably modeled by a bright Type
Ic supernova that occurred nearly contemporaneously with the GRB
\cite{bkp+02}. Moreover, this was the first burst for which the data
were of high enough quality to statistically demonstrate that the
afterglow propagated in a wind-stratified medium rather than a
constant-density ISM \cite{pbr+02}. I recently reviewed the pre-2003
evidence for the GRB-SN connection \cite{blo03}.

\label{sec:intro}

\section{GRB\,030329 and beyond}

While the case was rather strong for late-time emission bumps as being
due to underlying supernovae \cite{blo03}, the definitive direct
spectroscopic evidence was found only recently with the low-redshift
($z=0.17$; \cite{gpe+03}) GRB\,030329. For this burst, a number of
groups \cite{smg+03,cff+03,hsm+03,kdw+03} found a bump with supernovae
absorption features indicative of very high velocity ejecta. The close
evolutionary track of this apparent supernova (2003dh;
\cite{gmo+03})\index{2003dh} with SN\,1998bw should quell any
remaining doubt that long-duration GRBs are produced in the death of
massive stars.

The rare proximity of this event allowed for high-precision coverage
from the ground around the SN peak (e.g., Ref.~\cite{mgs+03,log+03a}).
Using optical/IR imaging at \hbox{CTIO 1.3m} to infer the
line--of--sight dust extinction, we inferred that the SN was roughly
0.3\,mag brighter at peak than the bright Type Ic 1998bw
\cite{bvdb+03}. Since the precise peak SN magnitude depends on the
line--of--sight extinction toward the GRB, it is imperative to infer
extinction from the early afterglow in future events. With new
dedicated robotic IR systems and space-based UV imaging, the broadband
spectral coverage (from 0.1 to 2 $\mu$m) of {\it Swift} afterglows
should routinely be used to infer line-of-sight extinction.

The peak brightness and early evolution of SN 2003dh, when compared to
1998bw, enabled estimates of the total mass of the ejecta (8
M$_\odot$), synthesized $^{56}$Ni mass (0.35 M$_\odot$), and mass of
the progenitor ($\sim$35 M$_\odot$) \cite{mdt+03}.  At this meeting, I
suggested that a great deal more could be learned with a concerted
long-term effort on HST to resolve and follow the evolution of the SN
as it faded below the integrated brightness of the host galaxy. A
measurement of the polarization of the light curve during the decay
phase would constrain the asphericity of the emission region
\cite{imn+98,wes99}. If the late-time evolution of SN 2003dh was indeed 
powered by radioactive decay of $^{56}$Co to $^{56}$Fe, then the time
history of the integrated luminosity would also determine the amount
of synthesized $^{56}$Ni (now only inferred by scaling 1998bw). To be
sure, leakage of $\gamma$-rays produced by radioactivity from the
expanding nebula would make the late-time decay steeper than a pure
$^{56}$Co exponential.

Though some of the details of SN 2003dh have not been flushed out
observationally, it seems clear that a bright core-collapsed supernova
and GRB 030329 occurred nearly simultaneously (to less than 1 week)
\cite{mgs+03,log+03a}. A pressing question is now whether a supernova 
accompanies all long-duration GRBs. On this point, a brief caution:
while bumps have been claimed in more than 10 events, a number of
these may be spurious due to improper subtraction of the host
light. Figure
\ref{fig:1} shows a compilation of the peak brightnesses of all
late-time bumps reported to-date in comparison with the total
magnitude of the respective host galaxy. Five bumps have been found to
peak within 0.3 magnitudes of hosts at $\sim$24--26\,mag,
uncomfortably close for photometric differencing at such faint levels.
\begin{figure}[tb]
\centering
\includegraphics[height=3.5in]{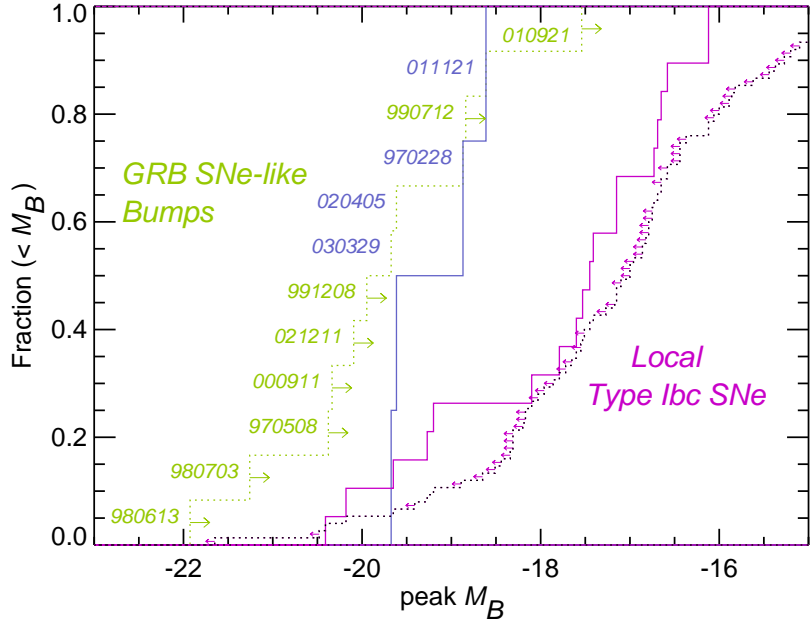}
\caption{Comparison of the peak magnitudes of GRB-SNe with
those of local Type Ib/Ic SNe. The solid cumulative histogram to the
left is for those GRBs with a believable detection of a SN bump; the
brightnesses of all other claimed GRB-SN peaks or reported
upper-limits are shown as a dotted histogram. The solid histogram to
the right is for those local Ibc SNe for which the SN was observed
before peak; all others were discovered after peak. The faintest
GRB-SN non-detection (010921) only probes the top
$\sim$40th-percentile of local Type Ib/Ic SNe.  It is clear that
current GRB-SNe population may have only revealed the tip of the
iceberg; plausibly, then, SNe could accompany all long-duration
GRBs. Compilation of local Ibc SNe from ref.~\cite{rbc+02}.}
\label{fig:2}       
\end{figure}

Of the bursts with redshift, I believe that only 030329, 020405,
970228, and 011121 can be legitimately claimed as true detections of
SN-like bumps related to a GRB. This statement is not intended to
imply that SNe do not accompany long-duration GRBs. Instead, the SN
peaks may simply have fallen below the detection threshold of the
instrument and/or the brightness of the host galaxy. How do these
non-detections compare with the expected brightnesses of GRB-SNe?
Even if all GRB-SNe are of type Ib/c, it is important to note that
there is no {\it a priori} theoretical reason to require that all
GRB-SNe should be identical: observed local Type Ib/c SNe show rather
diverse light curves, peak brightnesses and spectral evolution. Thus,
the simplest assumption is that the peak magnitude distribution of
GRB-SNe should follow the distribution of local Type Ib/c SNe. 

Figure \ref{fig:2} shows a comparison of the distribution of absolute
$B$-band peak magnitudes GRB-SNe (with known redshift) and local Ib/c
SNe. After correcting for the extinction due to dust in our Galaxy, I
have (simplistically) $k$-corrected the observed (or lower-limit) peak
magnitude to the restframe $B$-band using the observed redshift and
assuming a blackbody spectrum peaked in the restframe $V$-band. The
uncertainty introduced by this approximation on an individual $M_B$ is
likely less than 0.5\,mag. From the figure, it is clear that most
GRB-SNe detections (or limits) only probe the bright end of the local
Ib/c population. In fact, the deepest non-detection of a GRB-SN
(010921) only probes the brightest 40th percentile of local Ib/c
SNe. From Fig.~\ref{fig:2}, it is reasonable to conclude that the
observed supernovae related to GRBs have only probed the tip of the
iceberg of a plausible brightness distribution. A systematic survey of
late-time bumps from nearby ($z
\ale 0.5$) {\it Swift} bursts will test this hypothesis.

%
%
%
%



\printindex
\end{document}